\documentclass{osa-article}

\journal{oe}
\usepackage{lineno}
\usepackage{siunitx}
\begin{document}

\title{High-efficiency water-window x-ray generation from nanowire array targets irradiated with femtosecond laser pulses}

\author{Yinren Shou,\authormark{1} Defeng Kong,\authormark{1} Pengjie Wang,\authormark{1} Zhusong Mei,\authormark{1} Zhengxuan Cao,\authormark{1} Zhuo Pan,\authormark{1} Yunhui Li,\authormark{2} Shirui Xu,\authormark{1} Guijun Qi,\authormark{1} Shiyou Chen,\authormark{1} Jiarui Zhao,\authormark{1} Yanying Zhao,\authormark{1} Changbo Fu,\authormark{3} Wen Luo,\authormark{4} Guoqiang Zhang,\authormark{5} Xueqing Yan,\authormark{1,6,7} Wenjun Ma\authormark{1,6,7,*}}

\address{\authormark{1}State Key Laboratory of Nuclear Physics and Technology, and Key Laboratory of HEDP of the Ministry of Education, CAPT, Peking University, Beijing 100871, China\\
\authormark{2}Shenzhen Institutes of Advanced Technology, Chinese Academy of Sciences, Shenzhen 518055, China\\
\authormark{3}Key Laboratory of Nuclear Physics and Ion-Beam Application (MOE), Institute of Modern Physics, Fudan University, Shanghai 200433, China\\
\authormark{4}School of Nuclear Science and Technology, University of South China, Hengyang 421001, China\\
\authormark{5}Shanghai Institute of Applied Physics, Chinese Academy of Sciences, Shanghai 201800, China\\
\authormark{6}Beijing Laser Acceleration Innovation Center, Huairou, Beijing 101400, China\\
\authormark{7}Institute of Guangdong Laser Plasma Technology, Baiyun, Guangzhou 510540, China}

\email{\authormark{*}wenjun.ma@pku.edu.cn} %% email address is required



\begin{abstract}
We demonstrate the high-efficiency generation of water-window soft x-ray emissions from polyethylene nanowire array targets irradiated by femtosecond laser pulses at the intensity of \SI{4e19}{W/cm^2}. The experimental results indicate more than one order of magnitude enhancement of the water-window x-ray emissions from the nanowire array targets compared to the planar targets. The highest energy conversion efficiency from laser to water-window x-rays is measured as 0.5\%/sr, which comes from the targets with the longest nanowires. Supported by particle-in-cell simulations and atomic kinetic codes, the physics that leads to the high conversion efficiency is discussed.
\end{abstract}

%%%%%%%%%%%%%%%%%%%%%%%%%%  body  %%%%%%%%%%%%%%%%%%%%%%%%%%
\section{Introduction}
The developments of high-intensity laser technologies have opened the door to laser-driven light sources with photon energies from \SI{e-2}{eV} (THz radiation) to \SI{e7}{eV} (gamma-rays)\cite{mourou2019nobel,Renner2019}. Photons with the energies of $285\sim540$ \si{eV} ($\lambda \approx 2.3\sim4.4$ \si{nm}), also defined as the water-window (WW) x-rays, are ideal to image the inner structure of live cells in vivo with high spatial resolutions as they can penetrate the water while strongly be absorbed by the carbon atoms\cite{Anne2010,Kordel2020}. Laser-driven WW x-ray sources have the advantages of micrometer source sizes, ultrashort durations and high brightness\cite{albert2016}. Further improving the conversion efficiency from laser to WW x-rays is one key issue for the development of laser-driven laboratory WW x-ray microscopes, which can be achieved by utilizing novel targets. Besides regular planar solid targets\cite{Sheil2017,Arai2018,John2019}, plenty of novel targets such as liquid jet targets\cite{Groot2003}, gas-puff targets\cite{Muller2013,Wachulak2013}, carbon nanotube targets\cite{Nishikawa2004} and low-density foam targets\cite{Chakravarty2013,Hara2018} have been studied. Some of them showed appealing conversion efficiency from laser to WW x-rays.

As a kind of novel targets, nanowire array (NWA) targets have been used in high-power laser experiments for the enhanced generation of hot electrons\cite{Moreau2020}, energetic ions\cite{Dozieres2019,Ebert2020}, x-ray emissions\cite{Purvis2013,Hollinger2017} and neutrons\cite{Curtis2018}. Unlike planar targets where the laser pulses only heat the surface of the targets, the laser can penetrate into the NWA targets and interact with the side wall of the nanowires. As a result of such volumetric heating, the laser absorption of NWA targets is very high\cite{Habara2016}. Irradiated by ultrashort laser pulses with the intensity exceeding \SI{e18}{W/cm^2}, the nanowires can turn to high-energy-density plasmas with keV temperatures and near-solid densities. The relatively large volumes of the plasmas result in a long hydrodynamic cooling time of $T_h$. Here $T_h=L/C_s$, where $L$ is the plasma size and $C_s$ is the acoustic
velocity\cite{Hollinger2017}. As a result, the radiative cooling time $T_r$ is shorter than $T_h$, which means a large portion of plasma energy can be converted to x-ray emissions. A record conversion efficiency of 20\% from the laser to x-rays with the photon energy exceeding \SI{1}{keV} has been reported using Au NWA targets\cite{Hollinger2017}.

The wavelength of x-rays from the NWA targets is determined by the materials of the targets. Most NWA targets in previous studies were synthesized by electro-depositing metal atoms into anodized alumina oxide (AAO) membranes. The wavelengths of the emitted x-rays from the metal NWA targets are mainly in the range of $0.1-0.8$ \si{nm}. For example, Purvis et al. measured the x-rays in the wavelength range of $0.4-0.6$ \si{nm} from Au NWA targets\cite{Purvis2013}. X-ray emissions around \SI{0.15}{nm} were also reported from Ni and Co NWA targets\cite{Purvis2013,Bargsten2017}. For Si NWA targets, the wavelengths of emitted x-rays extended to $0.6-0.8$ \si{nm}\cite{Samsonova2019,Ebert2020}. All the above x-ray emissions were not in the WW region.

In this paper, we utilize polyethylene (PE) NWA targets prepared by the heat-extrusion method instead of the electro-deposition method to generate WW x-rays around \SI{3.5}{nm}. Experimental results show that the yield of WW x-rays from the PE NWA targets is more than one order of magnitude higher than that from the planar targets. The influence of the nanowires' lengths on x-ray emissions is experimentally investigated as well. We also perform a series of simulations to study the emission processes of WW x-rays from the NWA targets.

\section{Experimental set-up}
\begin{figure}[htbp]
  \centering
  \includegraphics[width=11cm]{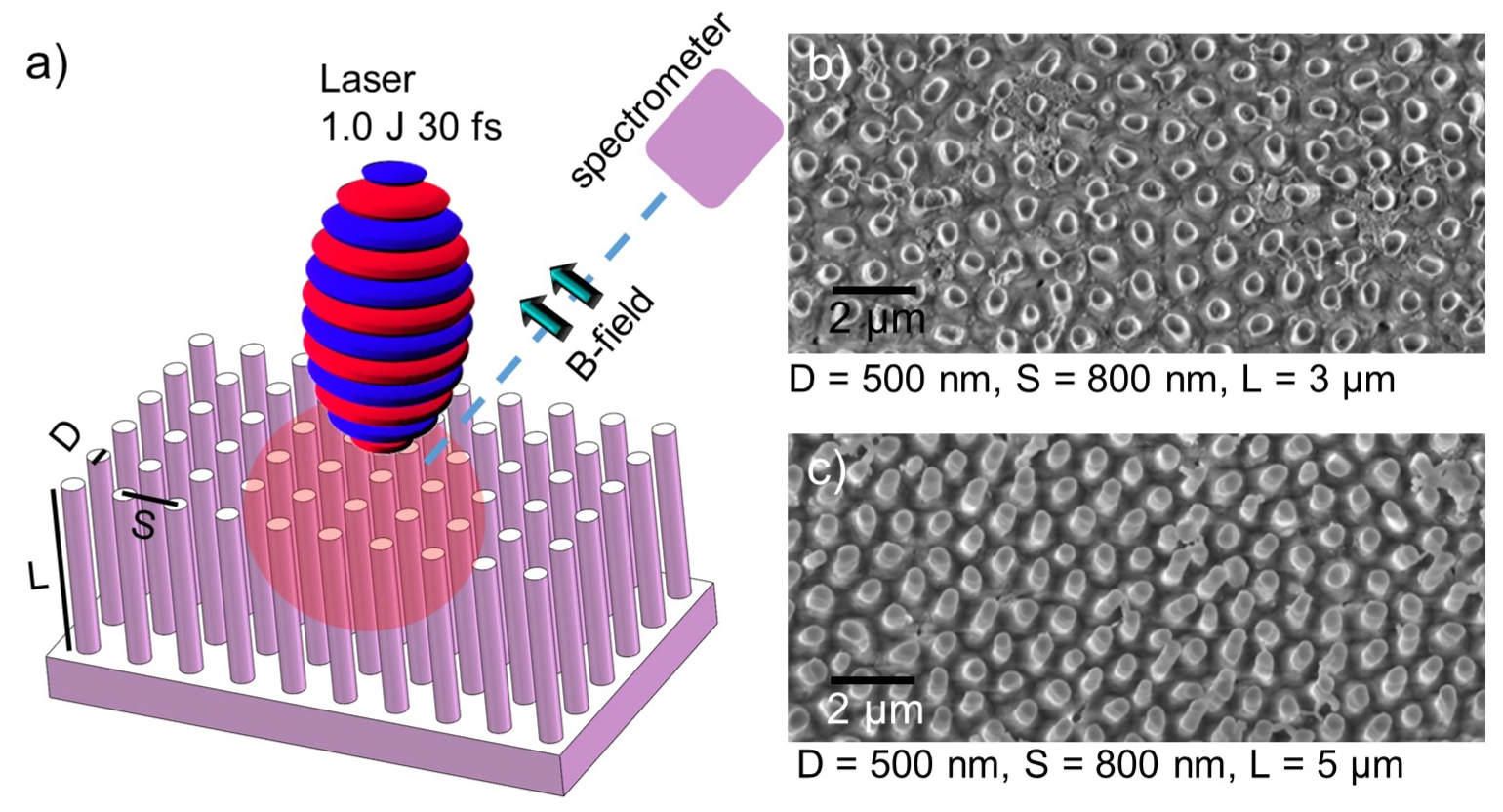}
\caption{The experimental set-up. a) A schematic diagram of the laser and the NWA targets. The laser incidence angle is  \SI{90}{\degree}. A flat-field grazing-incidence spectrometer is placed at \SI{45}{\degree} to the laser axis in the reflection direction to measure the x-ray emissions. b) and c) SEM images of the PE NWA targets with different lengths of nanowires.}\label{f1}
\end{figure}

The experiments were performed at Peking University utilizing the CLAPA laser system\cite{Geng2018}. As depicted in Fig. \ref{f1}a), a \SI{30}{fs} laser pulse with the central wavelength of \SI{800}{nm} was focused on the targets with \SI{90}{\degree} incident angle by an f/3 off-axis parabolic mirror (OAP). The CLAPA laser facility contains a cross-polarized wave (XPW) system to achieve a nanosecond contrast of $10^{10}$. To prevent the expansion of nanowires caused by the prepulses and amplified spontaneous emissions (ASE), we also employed a plasma mirror system to improve the laser contrast to $10^{-12}$ at \SI{40}{ps} before the main pulse. The on-target laser energy was about \SI{1.0}{J}. Considering the full width at half maximum (FWHM) spot size of \SI{4.0}{\micro m} $\times$ \SI{4.5}{\micro m} with an energy concentration ratio of 30\%, the peak intensity of the laser was \SI{4e19}{W/cm^2}. The corresponding maximum electric field was $E=$ \SI{1.6e13}{V/m}, which is strong enough to realize field ionization of carbon to C$^{6+}$\cite{ADK1986}.

Two kinds of PE NWA targets with the nanowires' diameters of \SI{200}{nm} and \SI{500}{nm} were used in the experiments. The nanowires' lengths varied from \SI{2}{\micro m} to \SI{10}{\micro m}. Under the nanowires, \SI{0.2}{mm} thick PE foils were employed as supporting substrates. Planar PE foils with the thickness of \SI{0.2}{mm} were also shot for comparison. The NWA targets were prepared by the heat-extrusion method with the following steps. First, we attached a PE sheet on a suitable AAO template composed of uniform parallel pores. Second, the PE-attached AAO templates were heated and mechanically compressed in vacuum so that the PE molecules were extruded into the pores of the template. Third, the NWA targets were obtained by dissolving the AAO templates in NaOH solutions. The scanning electron microscope (SEM) images of the PE NWA targets are displayed in Fig. \ref{f1}b) and c). As one can see, most nanowires stand straight and are isolated from each other. The diameters (D), interval spaces (S) and lengths (L) of the nanowires, as depicted in Fig. \ref{f1}a), are determined by the templates. In order to precisely place the targets on the laser focal spot, a motorized target positioning system was employed with an accuracy of \SI{2}{\micro m}\cite{Shou2019}.

The x-ray emissions from the NWA targets and the planar targets were measured by a flat-field grazing-incidence spectrometer at \SI{45}{^\circ} to the laser axis in the reflection direction. A 1200 lines/mm laminar-type soft x-ray diffraction grating (Shimadazu 03-005) was employed in the spectrometer. The WW x-ray spectra were recorded on-line using an x-ray charge-coupled device (CCD) camera (Andor DO940P-DN). The distance from the targets to the 55-\textrm{$\mu$}m-wide entrance slit of the spectrometer was \SI{0.98}{m}. Before the entrance slit, a magnet with a length of \SI{5}{cm} and a field strength of \SI{0.67}{T} was placed to deflect the ions and electrons generated from laser-plasma interaction.

\section{Results}
\begin{figure}[htbp]
  \centering
  \includegraphics[width=11cm]{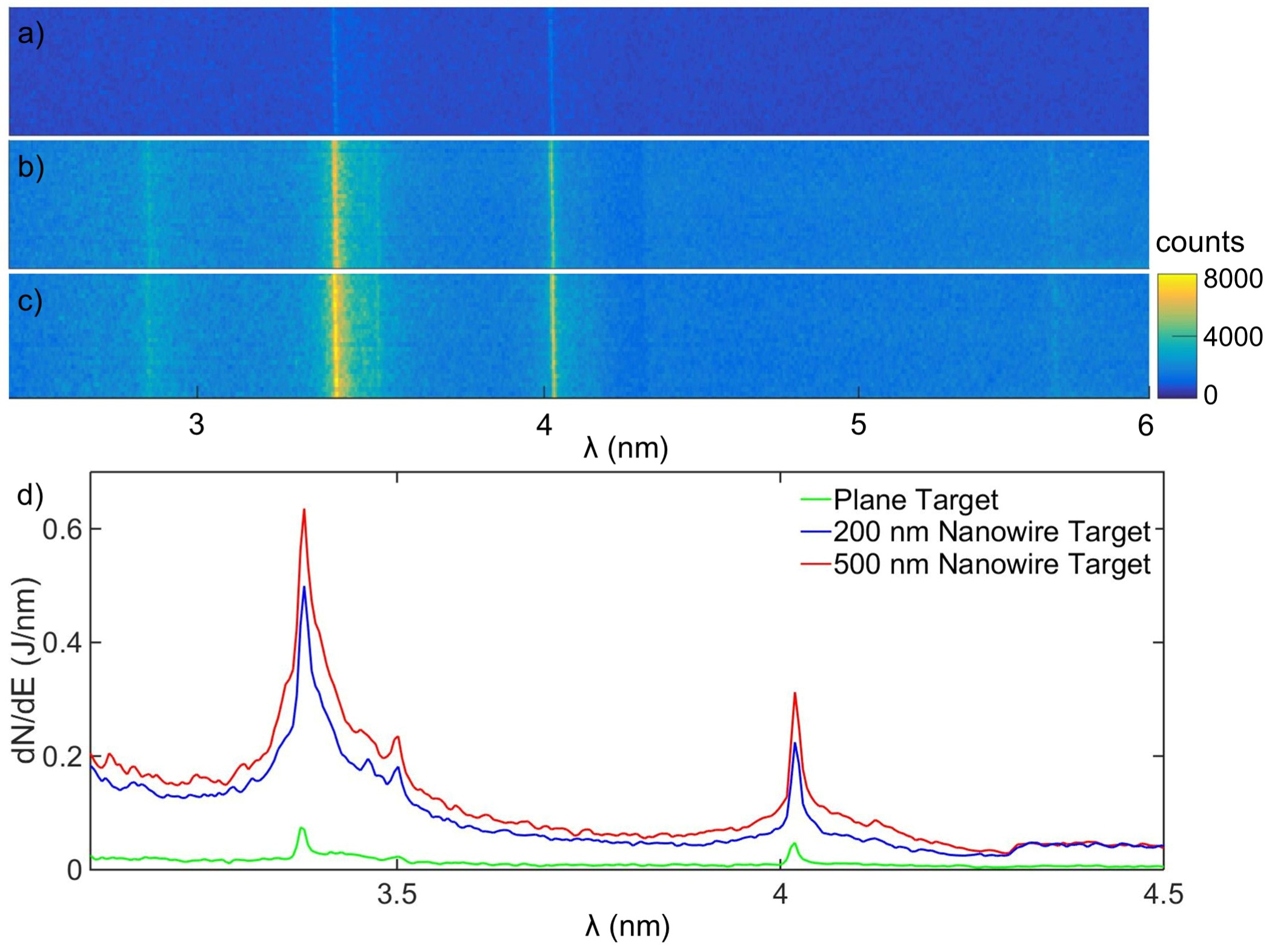}
  \caption{The WW x-ray spectra. a), b) and c) The raw data recorded by the x-ray CCD from planar targets, 200-nm-diameter NWA targets, and 500-nm-diameter NWA targets, respectively. The nanowires' lengths of the targets in b) and c) are both \SI{5}{\micro m}. d) The corresponding x-ray spectra of a) - c).}\label{f2}
\end{figure}

Figure \ref{f2}a) - c) display the typical raw data of the WW soft x-ray emissions measured by the flat-field spectrometer from the planar targets, NWA targets with the nanowires' diameters of \SI{200}{nm} and \SI{500}{nm}, respectively. Compared to the planar targets, a pronounced enhancement of x-ray emissions from the NWA targets can be observed. Taking into account the acceptance angle of the spectrometer, the diffraction efficiency of the flat-field grating\cite{Yamazaki1999,Dong2012} and the quantum efficiency of the x-ray CCD\cite{Andor}, we can obtain the x-ray spectra with the absolute spectral response as shown in Fig. \ref{f2}d). Here a 4$\pi$ solid angle of the x-ray radiation is assumed. The spectra of WW x-rays with the wavelength range of \SIrange{2.3}{3.1}{nm} are not included due to the extreme aberration of the spectrometer. As one can see, more than one order of magnitude enhancement on the x-ray emissions from both NWA targets over the planar targets is demonstrated. The 500-nm-diameter NWA targets have stronger emissions than that of the 200-nm-diameter NWA targets.

The measured WW x-rays are composed of continuum emissions and line emissions from carbon ions. The continuum emissions are from the recombination of free electrons with carbon ions and the bremsstrahlung radiation. At the laser intensity of \SI{4e19}{W/cm^2}, MeV relativistic electrons can be generated from the nanowires' surface and deposit their energy in nearby nanowires. So the bremsstrahlung radiation is significantly higher than that at lower intensities\cite{Sheil2017}, contributing to a stronger continuum emission. The line emissions riding over the strong continuum emission have the central wavelength of \SI{3.37}{nm} and \SI{4.03}{nm}, corresponding to the Ly$_\alpha$ and He$_\alpha$ emissions from C$^{5+}$ and C$^{4+}$, respectively. The K$_\alpha$ emission at \SI{4.48}{nm} from neutral carbon is not observed either from the planar or NWA targets. It indicates that most emissions are from ionized carbon ions instead of excited neutral ions. We speculate the relativistic driving laser leads to a quick field ionization of carbon atoms to C$^{4+}$ - C$^{6+}$ in the targets' surface, resulting in the strong Ly$_\alpha$ and He$_\alpha$ emissions as well as the suppression of K$_\alpha$ emissions.

\begin{figure}[htbp]
  \centering
  \includegraphics[width=8cm]{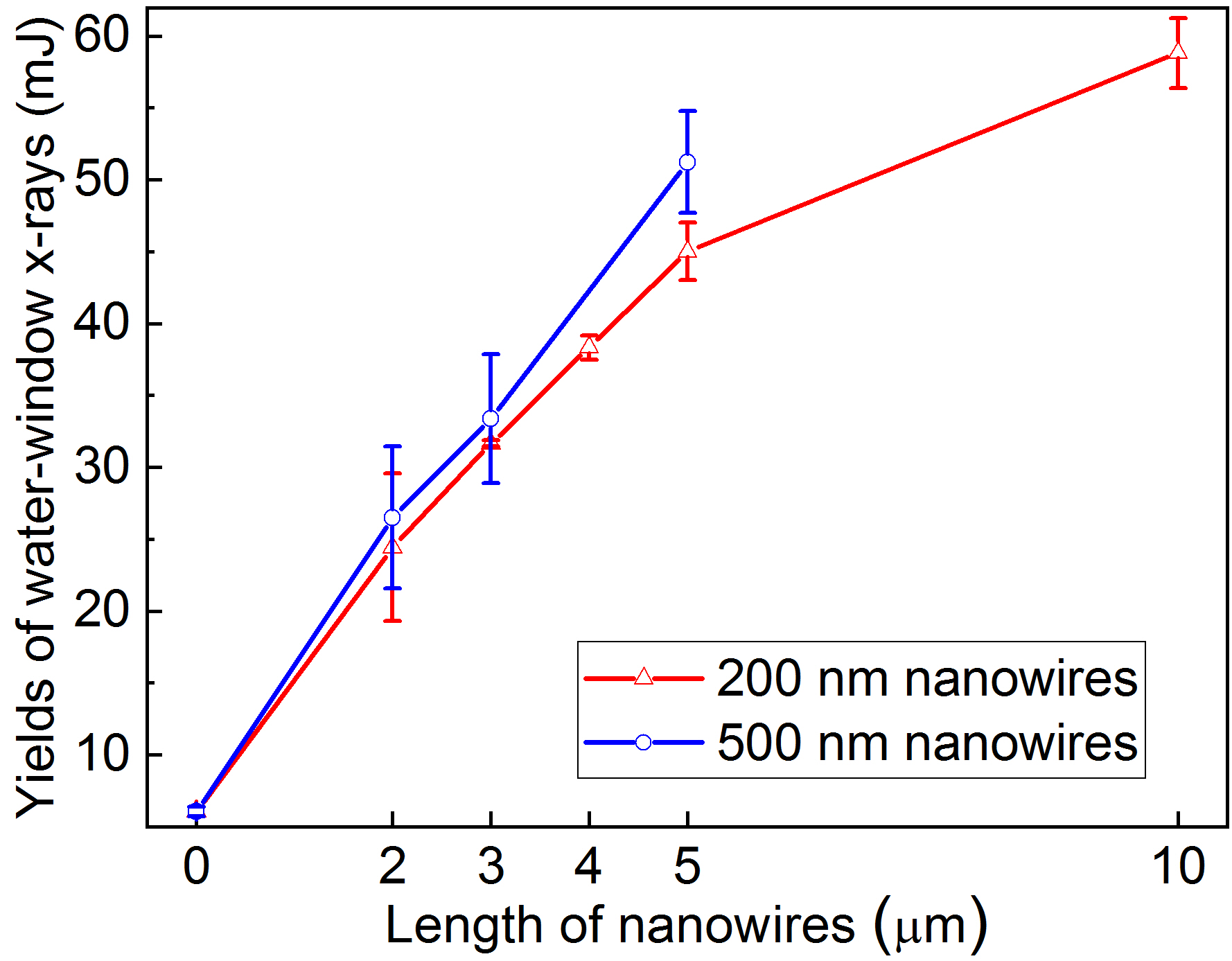}
\caption{Dependence of the WW x-ray yields on the lengths of nanowires for the 200-nm-diameter and 500-nm-diameter NWA targets. }\label{f3}
\end{figure}

We also studied the dependence of the nanowires' diameters and lengths on the yields of WW x-ray emissions. The yields are calculated from the integral of the WW x-ray spectra like Fig. \ref{f2}d). As shown in Fig. \ref{f3}, for the targets with the same length of nanowires, a slight enhancement of WW x-ray emissions can be observed for the 500-nm-diameter NWA targets compared to the 200-nm-diameter ones. The yields of WW x-rays rise with the increase of the nanowires' lengths. The conversion efficiency from laser to WW x-rays reaches 0.5\%/sr, or 6\% assuming a 4$\pi$ solid angle for the optimal 10-$\rm \mu m$-long NWA targets. When the lengths of nanowires exceed \SI{5}{\micro m}, the enhancement becomes less prominent. This tendency can be explained by the limited penetrating depth of the driven laser in the NWA targets, which was reported as \SI{6}{\micro m} in the previous work\cite{Bargsten2017}. Increasing the lengths of the nanowires from \SI{2}{\micro m} to \SI{5}{\micro m} can linearly enlarge the interaction volume, which eventually results in a higher laser absorption as well as stronger x-ray emissions. This effect becomes insignificant if the nanowires are longer than the penetrating depth of the laser.

\section{Discussion}
To investigate the electron heating and WW x-ray emission processes of the NWA targets, we first performed two-dimensional (2D) particle-in-cell (PIC) simulations utilizing the code EPOCH\cite{Arber2015}. The simulation box is \SI{6}{\micro m} $\times$ \SI{12}{\micro m} with a spatial resolution of \SI{2.5}{nm} $\times$ \SI{2.5}{nm}. A Gaussian laser pulse travels along the x axis from the left side of the simulation box with a central wavelength of \SI{800}{nm}. Its normalized vector potential, focal spot size and duration are $a_0=4$, \SI{4}{\micro m} and \SI{30}{fs}, respectively, the same as in the experiments. Here the normalized vector potential is $a_0=eE/m_ec\omega$, where $m_e$, $e$, $E$, $\omega$, and $c$ is the electron mass and charge, the peak electric field strength and angular frequency of laser, as well as the speed of light in vacuum, respectively\cite{macchi2013ion}. The NWA target is also set according to the experimental parameters. The PE plasma is composed of \SI{140}{n_c} electrons, \SI{20}{n_c} protons and \SI{20}{n_c} fully ionized carbons. Here the normalized density $\textrm{n}_\textrm{c}=$ \SI{1.7e21}{cm^{-3}}. The diameter, interval space and length of the nanowires are set as \SI{500}{nm}, \SI{800}{nm} and \SI{3}{\micro m}, respectively. The nanowires are in the x region of \SIrange{1}{4}{\micro m}. A \SI{1}{\micro m} thick PE plasma base is also set in the x region of \SIrange{4}{5}{\micro m}.

\begin{figure}[htbp]
  \centering
  \includegraphics[width=11cm]{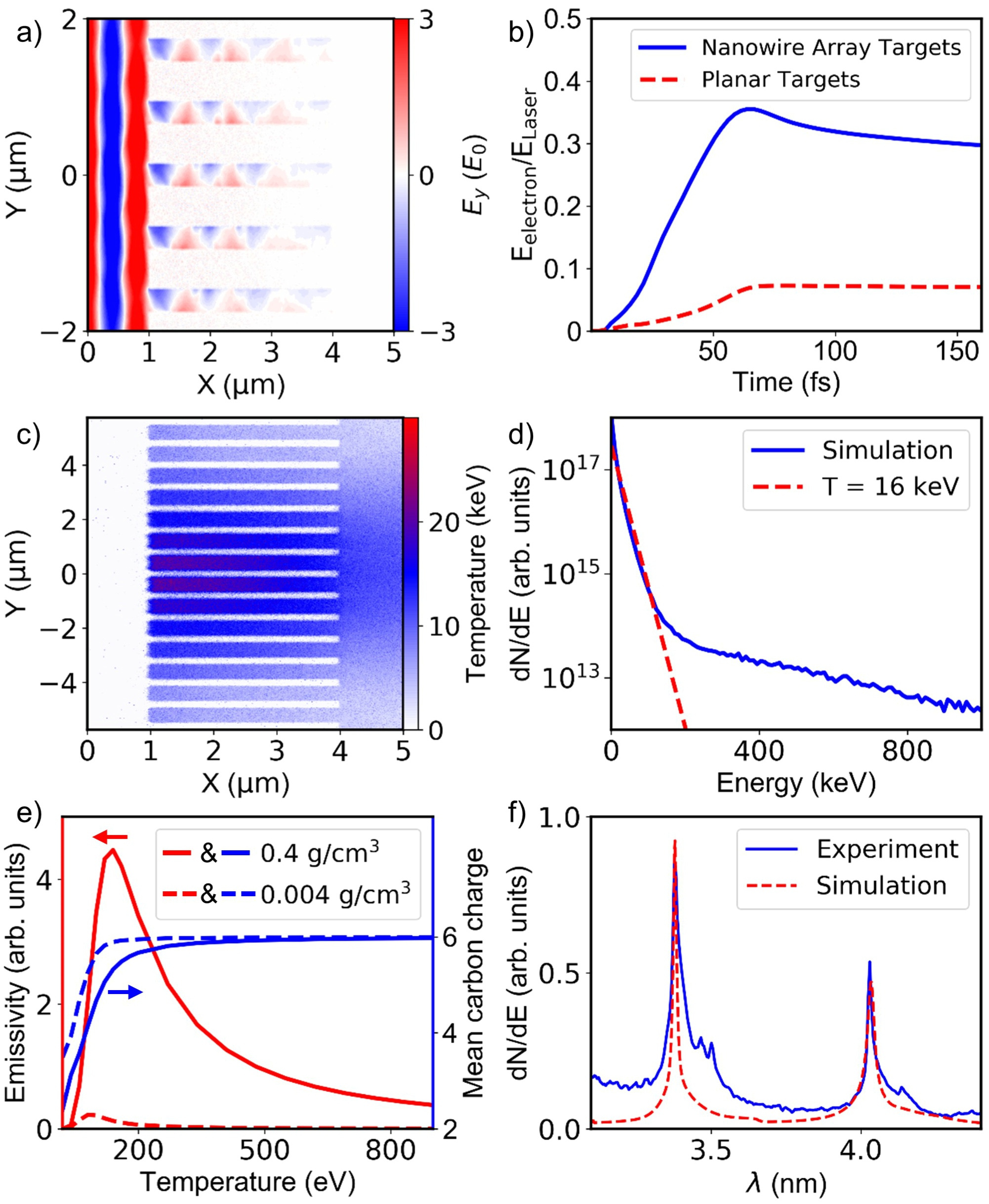}
\caption{The plasma heating and WW x-ray emission processes. PE NWA targets with the parameters of D = \SI{500}{nm}, S = \SI{800}{nm} and  L = \SI{3}{\micro m} are simulated. a) The spatial distribution of the electric field $E_y$ at the simulation time of $t=$ \SI{32}{fs}. $E_y$ is normalized by $E_0=m_ec\omega/e=$ \SI{4e12}{V/m}. b) The ratio of the electron energy $E_{\textrm{electron}}$ to the laser energy $E_{\textrm{laser}}$ in the simulation box for the planar and NWA targets. c) The distribution of the electron temperature at $t=$ \SI{100}{fs}. d) The energy spectrum of the electrons at $t=$ \SI{100}{fs}. The fitted bulk electron temperature is \SI{16}{keV} as shown by the red dashed line. e) Dependence of the bound-bound emissivity and mean charge state on the temperatures of carbon plasmas. Two mass densities of \SI{0.4}{g/cm^3} (solid lines) and \SI{0.004}{g/cm^3} (dashed lines) are calculated from the atomic kinetic code FLYCHK. f) The measured x-ray spectrum as well as the simulated one according to the estimated temporal evolution of the plasmas.}\label{f4}
\end{figure}

Figure \ref{f4}a) displays the spatial distribution of the electric field $E_y$ at the simulation time of $t=$ \SI{32}{fs}. The strong $E_y$ among the nanowires indicates that the driving laser can penetrate the NWA targets and interact with the side wall of the nanowires. As a result, the energy transferred to the electrons in the NWA targets can be enhanced by 4 times compared to that of the planar targets as shown in Fig. \ref{f4}b). Another remarkable phenomenon is that $E_y$ decays along the axis of nanowires from \SIrange{1}{4}{\micro m} in Fig. \ref{f4}a). It indicates a limited penetrating depth of the driving laser, consistent with the results in Fig. \ref{f3}.

The penetration of the laser in the NWA targets can result in the volumetric heating of the targets as depicted in Fig. \ref{f4}c). After the interaction of the laser, the distribution of electron temperature at the simulation time of \SI{100}{fs} indicates a volumetrically heated plasma. The electron temperature is slightly higher in the center of the laser spot due to the stronger laser field. We integrate the data in Fig. \ref{f4}c) to obtain the energy spectrum of the electrons as shown in Fig. \ref{f4}d). Considering that the number of high-energy electrons in the tail of the spectrum is very small compared to that of the bulk electrons, we focus on the low-energy but large-amount bulk electrons\cite{Sherlock2009, Rosmej2018}. The temperature of the bulk electrons in this simulation can be estimated as \SI{16}{keV} at $t=$ \SI{100}{fs} as shown by the red dashed line in Fig. \ref{f4}d).

The PIC codes can successfully simulate the laser heating process in the NWA targets. However, the emission process of WW x-ray photons is not included in the state-of-the-art PIC simulations. We utilized the atomic kinetic code FLYCHK\cite{Chung2005} to investigate the emission process of the laser heated plasmas. Calculations of the x-ray emissions from steady-state carbon plasmas were first performed to study the dependence of the bound-bound emissivity and mean charge state of carbon on the temperatures of the plasmas. As displayed in Fig. \ref{f4}e), one can see that the line emissions of carbon ions mainly happen at the temperature of 100s eV. At higher temperatures, most of the carbon atoms are ionized to C$^{6+}$, which suppresses the probability of line emissions from bound electrons. Figure \ref{f4}e) also indicates that the volumetric heating of solid density plasmas is another reason for the high conversion efficiency of NWA targets besides the efficient laser absorption. This can be reflected by the comparison of the bound-bound emissivity between plasmas with the mass densities of \SI{0.4}{g/cm^3} (the mean density of NWA targets) and \SI{0.004}{g/cm^3}. Compared to low-density plasmas, the high-density NWA plasma has a larger emissivity and a shorter radiative cooling time.

We also performed time-dependent simulations using FLYCHK to tentatively give the WW x-ray spectrum by considering the cooling process of the plasma. The key point is to get the temporal evolution of the plasma's temperature. We can not directly obtain it from PIC simulations because it requires too much computing resources if we want the simulate the whole process and meanwhile avoid numerical heating in the near-solid-density plasmas. So we use a semiquantitative model to estimate the evolution of the plasma's temperature. The PIC simulation indicates a hot plasma with the diameter of \SI{10}{\micro m} and the bulk electron temperature of \SI{16}{keV} is formed at $t=$ \SI{100}{fs}, which however is too hot to efficiently emit WW x-rays. Thereafter, a keV-electrons-driven heat dissipation process happens and results in a larger but colder plasma. This process is quite complicated due to the existence of the magnetic fields in the plasma\cite{Yang2021}. For simplicity, we only consider the process after the plasma's temperature drops below $T_0=$ \SI{600}{eV}, when efficient WW x-ray emissions happen according to Fig. \ref{f4}e). The corresponding plasma diameter $D_0$ can be estimated as $(16000/600)^{1/3}\times10\approx$ \SI{30}{\micro m}. Due to the high collision frequencies of electrons for such low-temperature plasma, a hydrodynamic cooling process will be dominant\cite{Purvis2013,Rolles2018}. Assuming the diameter of the hot plasma increases with the cooling time at the electrons' acoustic velocity of $C_s(t)=\sqrt{2T(t)/m_e}$, which is determined by the instantaneous temperature of the plasma, we have $T_0D_0^3=T(t)D^3(t)$. Here $D(t)=D_0+\int_0^t C_s(\tau)d\tau$ is the instantaneous diameter of the hot plasma. With the relationship of $dD(t)/dt=C_s(t)$, the derivative of $T(t)$ can be expressed as
\begin{equation}\label{e1}
  \frac{dT(t)}{dt}=-\frac{3C_s(t)T(t)}{D(t)}=-\frac{3\sqrt{2/m_e}}{D_0T_0^{1/3}}T(t)^{11/6}.
\end{equation}
Finally, the temporal evolution of the plasma temperature can be derived as
\begin{equation}\label{e2}
  T(t)=(T_0^{-5/6}+\frac{5}{2D_0}\frac{\sqrt{2/m_e}}{T_0^{1/3}}t)^{-6/5}.
\end{equation}
To include the effect of opacity, we also consider that the emitting area of the plasma surface increases as $D^2(t)$. Using the obtained plasma's temperature and the size of the emitting area as the input parameters, the time-dependent emissivity can be calculated by FLYCHK. By fixing the density of the NWA plasma as \SI{0.4}{g/cm^3}, the WW spectrum can be obtained by integrating the emissivity at different time as depicted in Fig. \ref{f4}f). The measured and simulated spectra fit each other quite well except the continuum part in the wavelength range of \SIrange{3.1}{3.8}{nm}, where the calculated spectrum is lower than the measured one. This difference may be due to the underestimate of the bremsstrahlung radiation as we only include the x-ray emissions in FLYCHK after the plasma is colder than \SI{600}{eV}, while electrons emit much more bremsstrahlung x-rays when the plasma is hotter.

\section{Conclusion}
In conclusion, we report the WW soft x-ray emissions from PE NWA targets irradiated by femtosecond laser pulses at relativistic intensity. The optimal conversion efficiency is measured as 0.5\%/sr, which is more than one order of magnitude higher compared to that of the planar targets. The yields of WW x-rays increase with the lengths of nanowires in the targets. We also perform simulations indicating that the high-efficiency generation of WW x-rays in NWA targets can be interpreted by the high laser absorption and the short radiative cooling time in the near-solid-density plasmas. Such a high-efficiency light source using NWA targets is quite suitable for laser-driven laboratory WW x-ray microscopes.

\section*{Funding}
National Grand Instrument Project (2019YFF01014402), NSFC innovation group project (11921006), and National Natural Science Foundation of China (Grant Nos. 11775010, 11535001, 61631001, 11905286).

\section*{Acknowledgments}
The PIC simulations were carried out in High-Performance Computing Platform of Peking University.

\section*{Disclosures}
The authors declare no conflict of interest.

%%%%%%%%%% If using BibTeX:
\bibliography{shou}

\end{document}